\documentclass[apjl]{emulateapj}

\usepackage{lscape}

\shorttitle{A parallax distance to V404 Cyg}
\shortauthors{Miller-Jones et al.}

\begin{document}

\title{The First Accurate Parallax Distance to a Black Hole}

\author{J. C. A. Miller-Jones\altaffilmark{1}}
\affil{NRAO Headquarters, 520 Edgemont Road,
  Charlottesville, VA 22902}
\email{jmiller@nrao.edu}

\author{P. G. Jonker\altaffilmark{2}}
\affil{SRON, Netherlands Institute for Space Research, 3584 CA,
  Utrecht, the Netherlands.}

\author{V. Dhawan, W. Brisken and M. P. Rupen}
\affil{NRAO Domenici Science Operations Center, 1003 Lopezville Road,
  Socorro, NM 87801}

\author{G. Nelemans}
\affil{Astrophysics Department, IMAPP, Radboud University,
  Toernooiveld 1, 6525 ED, Nijmegen, the Netherlands}

\and

\author{E. Gallo\altaffilmark{3}}
\affil{Massachusetts Institute of Technology, Kavli Institute for Astrophysics and Space Research, 70 Vassar St., Cambridge, MA 02139}

\altaffiltext{1}{Jansky Fellow.}
\altaffiltext{2}{Harvard-Smithsonian Center for Astrophysics,
    60 Garden Street, Cambridge, MA 02138}
\altaffiltext{3}{Hubble Fellow.}

\begin{abstract}
Using astrometric VLBI observations, we have determined the parallax
of the black hole X-ray binary V404 Cyg to be $0.418\pm0.024$\,mas,
corresponding to a distance of $2.39\pm0.14$\,kpc, significantly lower
than the previously accepted value.  This model-independent estimate
is the most accurate distance to a Galactic stellar-mass black hole
measured to date.  With this new distance, we confirm that the source
was not super-Eddington during its 1989 outburst.  The fitted distance
and proper motion imply that the black hole in this system likely
formed in a supernova, with the peculiar velocity being consistent
with a recoil (Blaauw) kick.  The size of the quiescent jets inferred
to exist in this system is $<1.4$\,AU at 22\,GHz.  Astrometric
observations of a larger sample of such systems would provide useful
insights into the formation and properties of accreting stellar-mass
black holes.
\end{abstract}


\section{Introduction}

Stellar-mass black holes in accreting X-ray binary systems provide
unique probes of the physical properties and formation mechanism of
black holes, as well as the physics of accretion and associated
outflow.  However, our understanding of black hole systems is hampered
by the fact that distances to X-ray binaries are relatively poorly
known, typically uncertain by 50\% or more \citep{Jon04}.  Standard
spectroscopic estimates are handicapped by the distortions induced in
the secondary by its compact partner, contamination by emission from
the accretion disc \citep{Rey08}, and uncertainties in estimates of
interstellar absorption.

Since these distance uncertainties are systematic rather than
random, they cannot be reduced by making more observations.  According
to \citet{Jon04}, these systematic uncertainties can artificially
enlarge the difference between the quiescent X-ray luminosities of
black hole and neutron star systems. If true this would undermine the
claimed evidence for an event horizon \citep{Nar97,Men99,Gar01}.
Accurate distances are also required to interpret the measured proper
motions of black hole X-ray binaries, by converting a measured proper
motion into a physical speed, from which we can derive the peculiar
velocity of the source \citep[e.g.,][]{Mir01,Mir02,Mir03,Dha07,Mil09}.
This can be used to place constraints on any velocity kick the black
hole might have received in its natal supernova, and hence on the
formation mechanism of the black hole \citep{Bra95,Wil05,Fra09}.

The only direct, model-independent method of measuring distances is
via trigonometric parallax, which has hitherto been impossible for
black hole X-ray binary systems, since they lie at distances of
several kpc, and hence require sub-milliarcsecond astrometry to
measure their parallaxes.  Very Long Baseline Interferometry (VLBI) at
radio wavelengths is currently the only technique available for such
high-precision astrometric measurements.  However, black hole X-ray
binaries spend the majority of their time in a faint quiescent state,
where they are often not detected in the radio band at current
sensitivities.  They become bright enough to be detected during
outbursts, but such outbursts are typically not sufficiently long or
frequent and do not sample the Earth's orbit well enough to make a
parallax measurement feasible.  During outbursts, the radio emission
is often resolved at the high angular resolutions required for
precision astrometry, making it difficult to determine the true
location of the binary system from epoch to epoch.  Lastly, many black
hole X-ray binaries are located in or close to the Galactic Plane,
such that the scatter-broadening along the line of sight makes
high-precision astrometry impossible in the centimeter waveband where
current VLBI arrays have the highest sensitivity.

Here we present High Sensitivity Array (HSA) observations of V404 Cyg,
the most luminous known black hole X-ray binary in quiescence.  With a
mass function of $6.08\pm0.06M_{\odot}$ \citep{Cas94}, the compact
object is a dynamically-confirmed black hole, accreting matter from a
K0 subgiant companion star \citep{Cas93}.  The system has a quiescent
radio flux density of 0.3\,mJy, with a flat spectrum indicative of a
self-absorbed compact jet \citep{Gal05}, although the jets are not
resolved at the angular resolution of the HSA \citep{Mil08}.  The
persistent, unresolved radio emission makes this source a good target
for astrometric observations to measure its parallax.

\section{Observations and results}
We observed V404 Cyg every 3 months over the course of one year, using
the HSA.  Two of the observations were made at the times of maximum
parallactic displacement in Right Ascension (R.A.), since the R.A.\
signal is larger, and, owing to the greater size of the array in the
east-west dimension, has smaller error bars than that in Declination.
We observed at a frequency of 8.4\,GHz, in dual circular polarization,
with an observing bandwidth of 64\,MHz per polarization.  To these
four HSA observations we added three archival datasets, two of which
have already been reported \citep{Mil08,Mil09}, and one in which the
source was previously not significantly detected \citep{Mio08}, but
knowing the astrometric parameters of the source, we were able to
detect it at the $5\sigma$ level.  A fourth archival epoch did not
detect the source to a $3\sigma$ limit of 0.36\,mJy\,beam$^{-1}$.  A
summary of the observations is listed in Table \ref{tab:obs}, and
shows the extent of the source variability.  In only one of the epochs
was the source brighter than the $3\sigma$ upper limit of this
archival dataset (in which none of the large dishes of the HSA were
available), making the non-detection consistent with the expected
source flux density.

\begin{figure}[t]
\epsscale{1.}
\plotone{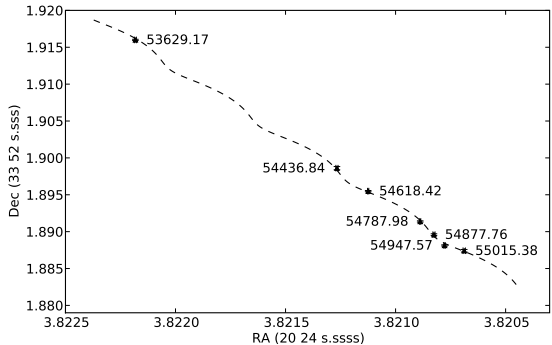}
\caption{Position of V404 Cyg at each of our 7 epochs of observation.
  Dashed line shows the best-fit proper motion and parallax.  Over
  time, the source moves south-west on the sky (from top left to
  bottom right of the figure).  Each position is labeled with the MJD
  of the mid-point of the observation.
  \label{fig:sky_positions}}
\end{figure}

In all cases, the observations were phase-referenced to a bright,
nearby calibrator from the International Celestial Reference Frame
(ICRF) source list, J\,2025+3343 \citep{Ma98}, located only
16.6$^{\prime}$ from the target source.  This implies that our
systematic errors, which scale with distance from the phase reference
source, should be relatively small ($\sim30$\,$\mu$as; see below).  We
observed in 3-min cycles, spending 1\,min on the phase reference
source and 2\,min on the target in each cycle.  To maximize
astrometric accuracy, all data below $23^{\circ}$ elevation were
discarded.  We also discarded data taken during the short timescale
flares seen in a number of the observations \citep[e.g.,][]{Mil08}, in
case the assumed variations in the jet power responsible for the
flaring events translated into positional offsets.  To enable us to
estimate the systematic errors affecting the astrometry, we
substituted every seventh scan on the target source with an
observation of a check source, the ICRF calibrator J2023+3153, located
$1^{\circ}.87$ away from the phase reference source.  Data were
reduced according to standard procedures within AIPS \citep{Gre03},
and the source position at each epoch was determined by fitting an
elliptical Gaussian to the source.  In no case did the source appear
to be resolved.  Since the first epoch of archival data assumed a
different position for the phase reference source, we corrected all
epochs to a common calibrator position of $20^{\rm h}25^{\rm m}10^{\rm
s}.8421056$, $33^{\circ}43^{\prime}00^{\prime\prime}.2144316$
(J\,2000) before fitting for the proper motion and parallax of V404
Cyg.  The measured positions are shown in
Figure~\ref{fig:sky_positions}, together with the best-fit parallax and
proper motion.

\citet{Pra06} used simulations to derive an approximate formula for
the systematic errors affecting various VLBI arrays.  For our
calibrator-target separation and source declination, the estimated
systematic errors (for a mean value of the wet zenith path delay) are
26\,$\mu$as in R.A.\ and 30\,$\mu$as in Dec.  The fitted positions of
the check source show an rms scatter of 13\,$\mu$as in R.A.\ and
31\,$\mu$as in Dec.  Since the check source is seven times further
away from the phase reference source than the target, this provides a
rigorous upper limit on the systematic errors in position.  As seen
from Table \ref{tab:obs}, our astrometric accuracy is limited by
signal-to-noise rather than systematic errors.  Nevertheless, the
systematic errors estimated from \citet{Pra06} were added to the
statistical errors in quadrature before using the singular value
decomposition (SVD) method \citep[as detailed in][]{Loi07} to fit for
a reference position and proper motion in R.A.\ and Dec., and for the
source parallax.  The best-fitting astrometric parameters, taking the
mid-point of the observations, MJD\,54322, as the reference date,
were:
\begin{displaymath}
\alpha_0 = 20^{\rm h}24^{\rm m}03^{\rm s}.821432\pm0^{\rm s}.000002\quad
{\rm (J\,2000)},
\end{displaymath}
\begin{displaymath}
\delta_0 =
33^{\circ}52^{\prime}01^{\prime\prime}.90134\pm0^{\prime\prime}.00005\quad
{\rm (J\,2000)},
\end{displaymath}
\begin{displaymath}
\mu_{\alpha}\cos\delta = -5.04\pm0.02{\rm \,mas\,yr}^{-1},
\end{displaymath}
\begin{displaymath}
\mu_{\delta} = -7.64\pm0.03{\rm \,mas\,yr}^{-1},
\end{displaymath}
\begin{displaymath}
\pi = 0.418\pm0.024{\rm \,mas}.
\end{displaymath}

\begin{figure}[t]
\epsscale{1.}
\plotone{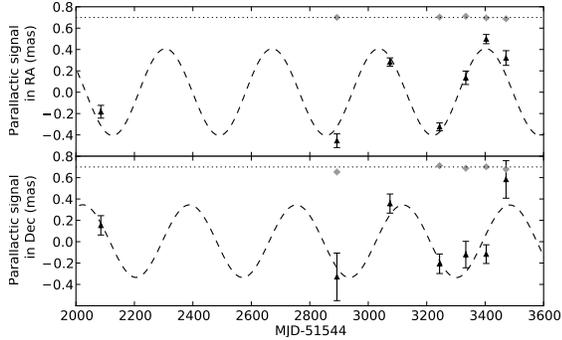}
\caption{Parallax signature of V404 Cyg in right ascension and
  declination.  The best-fit position and proper motion have been
  subtracted from the measured astrometric positions.  The grey points
  show the displacement of the check source from its weighted mean
  position (dotted line), shifted by 0.7\,mas for clarity, and scaled
  by a factor 1/7 to correct for its relative distance from the phase
  reference source.  This illustrates the scale of the systematic
  errors.\label{fig:par_signal}}
\end{figure}

The fitted parallax signal, with the best-fitting proper motion
removed, is shown in Figure~\ref{fig:par_signal}.  The reduced-$\chi^2$
value of the fit is 3.2, which, with 9 degrees of freedom, implies a
$>95$\%-confidence result.  A bootstrap data-stripping analysis, as
detailed by \citet{Cha09}, confirmed the validity of these results,
giving median values differing by less than $1\sigma$ from those found
by the SVD fit, albeit with slightly larger error bars owing to the
data stripping.  Our fitted parallax corresponds to a source distance
of $2.39\pm0.14$\,kpc.

\section{Discussion}

\subsection{Extinction-based distances}
Our fitted source distance is significantly closer than the best
previous distance estimate of $4.0^{+2.0}_{-1.2}$\,kpc \citep{Jon04}.
However, that estimate was derived using an assumed interstellar
extinction of $A_V=3.3$, and did not take into account uncertainties
in the extinction.  \citet{Hyn09} fitted the full multiwavelength
spectral energy distribution with a template spectrum for a K0 IV star
and found a formal best fit reddening of $A_V=4.0$.  Indeed, we find
that increasing $A_V$ to close to 4.0 brings the estimate of
\citet{Jon04} down into agreement with our results.  This highlights
the importance of using accurate extinction values when estimating
source distances.  Given the typical uncertainties in the extinction
towards black hole soft X-ray transients \citep{Jon04}, we conclude
that this issue is likely to affect the majority of such systems,
rendering their distance estimates also uncertain.

\subsection{Source luminosity}
Our measured distance is significantly closer than the 3.5--4.0\,kpc
commonly assumed for the source
\citep[e.g.,][]{Jon04,Gal05,Bra07,Cor08}, reducing its luminosity by a
factor of 2.2--2.8.  The quiescent 0.3--10\,keV unabsorbed flux of
$1.08\times10^{-12}$\,erg\,cm$^{-2}$\,s$^{-1}$, derived using a
power-law model for the X-ray spectrum \citep{Bra07}, implies a
quiescent luminosity of $\sim 7\times10^{32}$\,erg\,s$^{-1}$.
\citet{Mak89} measured a maximum flux of 17\,Crab with {\it Ginga}
during the 1989 outburst of V404 Cyg, implying a 1--70\,keV luminosity
of $7.0\times10^{38}$\,erg\,s$^{-1}$, which is of order $0.5L_{\rm
Edd}$ for the black hole mass of $12\pm2M_{\odot}$ derived for V404
Cyg \citep{Sha94}.  Thus the system was not super-Eddington during the
1989 outburst.  Furthermore, in reducing the source radio and X-ray
luminosities, our new distance will reduce the scatter in the
radio/X-ray correlation \citep{Gal03}, bringing the points measured
for V404 Cyg into better alignment with those of GX 339-4.

\subsection{Peculiar velocity}
With a more accurate distance and proper motion for the source, and
the updated Galactic rotation parameters ($R_{\odot}=8.4$\,kpc,
$\Theta_{\odot}=254$\,km\,s$^{-1}$) given by \citet[][]{Rei09}
\citep[but note][]{McM09}, we can revisit the analysis of
\citet{Mil09} to derive a more accurate peculiar velocity of
$39.9\pm5.5$\,km\,s$^{-1}$.  This error bar now includes the
uncertainties in both the distance and in all three space velocity
components.  This new value is significantly lower than the
best-fitting peculiar velocity of 64\,km\,s$^{-1}$ derived for a
4\,kpc distance \citep{Mil09}, and can easily be achieved via a Blaauw
kick, such that no asymmetric supernova kick is required.  The
component of the peculiar velocity in the Galactic Plane is
39.6\,km\,s$^{-1}$, which, compared to the expected velocity
dispersion in the Galactic Plane of 18.9\,km\,s$^{-1}$\citep{Mig00}
for the likely F0-F5 progenitor of the donor star, is a $2.1\sigma$
result, implying a probability of only 0.038 that the peculiar velocity
is a result of Galactic velocity dispersion \citep[for more
  details, see][]{Mil09}.  We therefore find it most probable that the
peculiar velocity arises from a natal supernova kick, the magnitude of
which is consistent with recoil due to mass loss (a Blaauw kick), with
any additional asymmetric kick being small.

\subsection{Size constraints}

\citet{Mil09} placed an upper limit of 1.3\,mas on the source size,
corresponding to a physical size of $<3$\,AU at our new distance.
While we did not resolve the source, our highest-resolution data,
taken with a global VLBI array at 22\,GHz, constrain the source size
to $<0.6$\,mas, a physical size of $<1.4$\,AU.  We can therefore place
a lower limit of $10^6$\,K on the brightness temperature of the
source.  From the high brightness temperature, the observed flat radio
spectrum \citep{Gal05} characteristic of a steady, partially self-absorbed
conical outflow \citep{Bla79}, and the location of the source on the
radio/X-ray correlation of \citet{Gal03}, at the high-luminosity end
of which jets have been directly resolved \citep{Sti01}, we infer that
the observed radio emission is likely to arise from compact, steady
synchrotron-emitting jets.

Our upper limit on the jet size can be compared to that of the
resolved jets in Cygnus X-1 \citep{Sti01}, with an angular size of
15\,mas at 8.4\,GHz.  Since the jet size scales inversely with
observing frequency \citep{Bla79}, this implies 5\,mas at 22\,GHz, for
a physical scale of $10(d/2{\rm kpc})$\,AU.  Since \citet{Hei06} found
the jet length should scale as $L_{\nu}^{8/17}$ where $L_{\nu}$ is the
jet luminosity, then the difference in radio luminosities of the two
sources suggests a size scale of $\sim1.5$\,AU for V404 Cyg at 22\,GHz.
Thus we are beginning to probe the angular scales on which the jets
may be resolved.

The scatter in the residuals after our parallax fit (0.10\,mas in
R.A.\ and 0.14\,mas in Dec.) can also help constrain the size and
stability of the jets, since any motion of unresolved emitting knots
along the jet will change the brightness distribution and hence the
measured centroid position.  In particular, a comparison of source
positions measured during flares with those from the adjacent
non-flaring data from which the proper motion and parallax were
derived will constrain the changes in jet morphology during flaring
events.  If the magnitude of the flares correlates with jet power and
hence jet size, we might expect brighter flares to show the largest
positional shifts.  For our brightest flare, in which the flux density
increased by a factor of 3, the $3\sigma$ upper limit to the shift in
the centroid position is 0.3\,mas.  This constrains the emitting
region during the flare to be at an angular separation of $<0.45$\,mas
(a physical separation of $<1$\,AU) from the surface in the steady jet
where the optical depth is unity.  While such small positional shifts
imply that minor flaring events should not destroy the parallax
signal, if they can be better constrained during larger flares or
with a higher-sensitivity array, such positional shifts could in
future provide a method for determining the size scale of quiescent
jets even in cases where they cannot be directly resolved.

\subsection{Wider implications}
These observations demonstrate that it is feasible to measure a
parallax distance to a sufficiently bright quiescent black hole X-ray
binary.  We have shown that the radio jets inferred to exist in the
quiescent state of X-ray binaries do not hinder the astrometry
sufficiently to render a parallax measurement impossible.  The advent
of the EVLA and the ongoing VLBA sensitivity upgrade will allow us to
extend this method to a number of systems with fainter quiescent jets,
as well as opening up the possibility of detecting gyrosynchrotron
emission from low-mass donor stars \citep[e.g.,][]{Gud02}.  With
accurate distances to a sample of quiescent black hole and neutron
star systems, it will be possible to quantitatively assess the claim
that black holes are less luminous than neutron star systems
\citep{Nar97,Men99,Gar01}.  Accurate outburst luminosities for a
sample of black holes will help quantify the factor by which such
systems can exceed their Eddington luminosities, with implications for
the nature of Ultraluminous X-ray sources (ULXs) in external galaxies.

With a sample of black hole proper motions and parallax distances, it
will be possible to derive accurate peculiar velocities to probe the
formation mechanisms of black holes of different masses.  The three
main channels are direct collapse (giving no peculiar velocity), or
via a supernova and fallback onto the proto-neutron star, with either
a relatively small recoil kick for a symmetric supernova, or a
potentially large kick from an asymmetric explosion.  In the fallback
case it is currently unclear whether and how the velocity scales with
the amount of matter falling back on the neutron star.  A sample of
black hole peculiar velocities would provide much-needed observational
constraints, and help to determine whether there is a mass
cutoff between black holes forming via direct collapse and those which
form via a supernova and fallback \citep{Fry01}.  Finally, one of the
methods currently employed to measure black hole spins, via fitting of
the X-ray spectrum during the thermal-dominant state of the source
during outburst \citep{McC06} requires accurate distances.  Better
distance constraints will allow for improved comparisons with spins
derived via fitting the general relativistic distortions in disk
reflection lines \citep[e.g.,][]{MilJM09}, providing further insights
into the generation and distribution of black hole spins.

\section{Conclusions}
We have used HSA observations to measure the most accurate distance to
a black hole to date, via the method of trigonometric parallax.  The
distance of V404 Cyg is $2.39\pm0.14$\,kpc, significantly lower than
the previous best estimate of the source distance, and demonstrating
that uncertainties in the interstellar extinction can introduce
significant errors into distance estimates.  With the new distance, we
can determine that V404 Cyg did not exceed its Eddington luminosity
during the 1989 outburst, and that it is likely to have formed in a
natal supernova, with the peculiar velocity being consistent with
recoil from the ejected mass, and any asymmetric kick being small.  We
have further constrained the size of the quiescent jets in the source
to $<1.4$\,AU.  These observations demonstrate the feasibility of
measuring parallax distances to black hole X-ray binaries using VLBI
techniques, which could, with a larger sample of systems, firm up
evidence for black hole event horizons, and provide new insights into
black hole spins, the mechanism of black hole formation, and the
nature of ULXs.

\acknowledgments 
J.C.A.M.-J.\ is a Jansky Fellow of the National Radio Astronomy
Observatory (NRAO).  E.G.\ is funded by NASA through a Hubble
Fellowship grant from the Space Telescope Science Institute, which is
operated by the Association of Universities for Research in Astronomy,
Inc., under NASA contract no. NAS5-26555.  P.G.J.\ acknowledges a VIDI
grant from the Netherlands Organization for Scientific Research.  The
GBT, VLA and VLBA are facilities of the NRAO which is operated by
Associated Universities, Inc., under cooperative agreement with the
National Science Foundation.  The European VLBI Network is a joint
facility of European, Chinese, South African and other radio astronomy
institutes funded by their national research councils.  This research
has made use of NASA's Astrophysics Data System.

{\it Facilities:} \facility{VLBA}, \facility{GBT}, \facility{VLA}.

\clearpage 

\begin{landscape}
\begin{deluxetable}{llllccccc}
%\tabletypesize{\scriptsize}
%\rotate
\tablecaption{Astrometric observations of V404 Cyg\label{tab:obs}}
\tablewidth{0pt}
\tablehead{
\colhead{Project} & \colhead{Date} & \colhead{MJD\tablenotemark{a}} & \colhead{Array} &
\colhead{Frequency} & \colhead{Bandwidth} & \colhead{Flux density\tablenotemark{b}} & \colhead{R.A.\tablenotemark{c}} &
\colhead{Dec.\tablenotemark{c}}\\
& & & & (GHz) & (MHz) & (mJy\,beam$^{-1}$) & $20^{\rm h}24^{\rm m}$ & $33^{\circ}52^{\prime}$}
\startdata
BM117 & 1999 Apr 10 & 51278.60(25) & VLBA & 15.353 & 32 & $<0.36$\tablenotemark{d} & - & - \\
BM231 & 2005 Sep 16 & 53629.17(21) & VLBA, VLA & 15.365 & 32 & $0.33\pm0.05$ & $03^{\rm s}.822182(5)$ & $01^{\prime\prime}.91599(9)$\\
BG168 & 2007 Dec 02 & 54436.84(09) & VLBA, GBT, VLA & 8.421 & 32 & $0.28\pm0.03$ & $03^{\rm s}.821266(5)$ & $01^{\prime\prime}.89861(22)$\\ 
GM064 & 2008 Jun 01 & 54618.42(25) & VLBA, GBT, VLA, EVN\tablenotemark{e} & 22.220 & 64 & $0.13\pm0.03$ & $03^{\rm s}.821125(2)$ & $01^{\prime\prime}.89549(8)$\\
BM290 & 2008 Nov 17 & 54787.98(09) & VLBA, GBT & 8.408 & 64 & $0.44\pm0.03$ & $03^{\rm s}.820888(3)$ & $01^{\prime\prime}.89138(9)$\\
BM290 & 2008 Feb 15 & 54877.76(11) & VLBA, GBT, VLA & 8.408 & 64 & $0.17\pm0.02$ & $03^{\rm s}.820826(5)$ & $01^{\prime\prime}.88959(12)$\\
BM290 & 2009 Apr 26 & 54947.57(11) & VLBA, GBT, VLA & 8.408 & 64 & $0.33\pm0.02$ & $03^{\rm s}.820777(3)$ & $01^{\prime\prime}.88813(8)$\\
BM290 & 2009 Jul 03 & 55015.38(10) & VLBA, GBT, VLA & 8.408 & 64 & $0.13\pm0.02$ & $03^{\rm s}.820688(5)$ & $01^{\prime\prime}.88741(17)$\\
\enddata
\tablenotetext{a}{Specified Modified Julian Dates (MJDs) are for the
  mid-point of the observation, with the quoted uncertainties reflecting the
  observation duration}
\tablenotetext{b}{Flux density measured after discarding short-timescale flaring events}
\tablenotetext{c}{All positions are specified in J\,2000 co-ordinates}
\tablenotetext{d}{Source not significantly detected at this epoch; $3\sigma$ upper limit given on flux density}
\tablenotetext{e}{Data taken with a global VLBI array, comprising the
  VLBA, GBT, phased VLA, seven dishes of the EVN (Cm, Eb, Jb2, Mc, Mh,
  Nt, Tr), and the Robledo antenna from the DSN}
\end{deluxetable}
\clearpage
\end{landscape}

\end{document}